\documentclass[10pt, conference, compsocconf]{IEEEtran}
\IEEEoverridecommandlockouts
\usepackage{cite}
\usepackage{amsmath,amssymb,amsfonts}
\usepackage{algorithmic}
\usepackage{graphicx}
\usepackage{textcomp}
\usepackage{xcolor}

\begin{document}
    
    % describe the protocol for control access (blockchain part)
    %  Related work
    %    -   traditional architectures commonly employed in the IoT access:
    %       XACML, OAuth, UMA
    %       Dependency of a central entity

\title{Blockchain-based Bidirectional Updates on Fine-grained Medical Data}

\author{
    \IEEEauthorblockN{%
        Chunmiao Li\textsuperscript{1,3},
        Yang Cao\textsuperscript{2},
        Zhenjiang Hu\textsuperscript{1,3,4},
        Masatoshi Yoshikawa\textsuperscript{2}
    \vspace{1.5ex}
    \IEEEauthorblockA{%
        \textsuperscript{1}\,National Institute of Informatics, Japan\qquad
        \textsuperscript{2}\,Kyoto University, Japan \\
        \textsuperscript{3}\,SOKENDAI (The Graduate University for Advanced Studies), Japan\qquad
        \textsuperscript{4}\,University of Tokyo, Japan
    }
%    \vspace{1ex}
%    \IEEEauthorblockA{%
%        Email:\
%        \textsuperscript{1}\,chunmiaoli1993@nii.ac.jp,
%        \textsuperscript{2}\,yang@i.kyoto-u.ac.jp,
%        \textsuperscript{3}\,hu@nii.ac.jp,
%        \textsuperscript{4}\,yoshikawa@i.kyoto-u.ac.jp
%    }
}
}

\maketitle

%regulate the use of a, an, the
\begin{abstract}
Electronic medical data sharing between stakeholders, such as patients, doctors, and researchers, can promote more effective medical treatment collaboratively. These sensitive and private data should only be accessed by authorized users.  Given a total medical data, users may care about parts of them and other unrelated information might interfere with the user-interested data search and increase the risk of exposure. Besides accessing these data, users may want to update them and propagate to other sharing peers so that all peers keep identical data after each update. To satisfy these requirements, in this paper we propose a medical data sharing architecture that addresses the permission control using smart contracts on blockchain and splits data into fined-grained pieces shared with different peers then synchronize full data and these pieces with bidirectional transformations. Medical data reside on each user's local database and permission-related data are stored on smart contracts. Only all peers have gained the newest shared data after updates can they start to do next operations on it, which are enforced by smart contracts. Blockchain-based immutable shared ledge enables users to trace data updates history. This paper can provide a new perspective to view full medical data as different slices to be shared with various peers but consistency after updates between them are still promised, which can protect privacy and improve data search efficiency.

\end{abstract}

\begin{IEEEkeywords}
medical data, blockchain, update, bidirectional transformations
\end{IEEEkeywords}

\section{Introduction}
Now a lot of medical data are digitalized so as to be stored and accessed conveniently. A medical record is produced after a patient goes to see a doctor and often resides on the hospital's database. Medical records usually contain highly sensitive information about patient privacy. HIPPA Privacy Rule  \cite{centers2004hipaa} in the U.S. regulates the use and disclosure of personally identifiable health information to protect patients' privacy. However, it is hard to make sure that all medical institutes would follow these rules because they may expose patient privacy for profit. Hence, transparency in medical data management system is important. Moreover, patients might visit different hospitals and leave their records scattered \cite{zhang2016secure} in different places, so that it is hard for them to manage records efficiently. Patients should better to be provided with a platform to manage and review their historical medical data that are from different hospitals in case of exposure or being tampered. In addition to providing data to doctors, 
sharing medical data could also benefit other stakeholders such as researchers or policy makers.
For example, researchers can identify public health risks and then develop better treatment by analyzing large-scale medical data \cite{office2015report}. 
%As presented in \cite{chung2018using},  patients and experts can exchange information to develop better plans to satisfy individual routines. 

% %patient understand and trace his medical records
% %doctor revise treatment plan in terms of feedback
% %researcher conduct census analyzation and provide better measures

To protect patients' data from being exposed or tampered, shared medical data could reside in encrypted formats on a trusted cloud storage server and can only be accessed by authorized users. However, in that case, centralized access control might lead to a single point of failure and become the bottleneck of the sharing system. Some decentralized medical data sharing systems \cite{azaria2016medrec,fan2018medblock,xia2017bbds} have been proposed to manage authentication based on blockchain\cite{nakamoto2008bitcoin} technology, which achieves consensus among distributed nodes. The access control logic of medical data on smart contracts\cite{azaria2016medrec} or Chaincode \cite{dubovitskaya2017secure} can be used as a criterion to judge whether a user can be allowed to access medical data. Only a user satisfies that permission information can his access be agreed by the majority of nodes, which means he is authorized from the blockchain side.

% and their access process will be recorded on the blockchain.

Still, we identified a limitation on current works. 
%Users might be overwhelmed by a full medical record since they tend to have a unique focus on it. 
Consider that a full medical record can have many attributes, while different parties may be only interested in specific parts of them. For example, given a medical record, researchers are interested in the attribute of the mechanism of medicine action, whereas patients care more about the medicine dosage standard attribute.  Moreover, additional but unnecessary information might influence or even mislead users' judgment. Imagine that doctors may add some symptom description on records which might put patients in confusion and fear \cite{delbanco2010open}. Besides, treatment steps are exclusive to a hospital and should not be directly accessed by other users.

To fill this gap, we propose an idea that a full medical data (i.e., data source) could be split into lots of smaller pieces (i.e., data views). From the perspective of relational databases, these kinds of smaller pieces can be seen as view tables derived by querying a few but not all attributes on the base (source) table. A user can share different fine-grained data pieces with different users based on predefined protocols. Imagine a doctor can share dosage usage with a patient and medicine mechanism with a researcher respectively. In this way, only user-concerned data are exposed to them which can avoid additional data interference and protect private proprietary data of data providers from being leaked. 

However, if we adopt this idea, we have to dispose of two issues as follow.

Firstly, we need to handle the synchronization between source and multiple views after updates on those fine-grained views occur. Consider this scenario: a researcher updates the medicine mechanism on shared data with a doctor. Note here the shared data is actually a view from the full medical data (source) on the doctor side. Thus the doctor needs to synchronize this change on view to create a new source. To solve this problem, we apply bidirectional transformations \cite{hu2014validity} to synchronize them after updates on either one side. For example, we can invoke \emph{put} direction of a  BX program to reflect modifications on shared data in the full data and \emph{get} to produce shared data from full data. Because different views produced from the same source might have overlapping data. After the updates on shared data are reflected to doctor's source, the doctor still has to judge whether he needs to modify the shared data with patients by reproducing a new view.

%Current works mostly only consider uni-directional updates on shared data, which means that these updates are conducted by data providers and then produce modified data to other users. They did not offer the synchronization scheme between source and views on the same user side.

%read, add, delete shared data belong to data management.
%Since the shared data just reside in each peer's local database, read the shared data is easily promised.

%For authorization
Secondly, we need to conduct access control on fine-grained view data. Existing blockchain-based access control solutions, such as \cite{azaria2016medrec}, focused on permission control on the full record. However, we can adapt it to work with the fine-grained views. For example, we can encode permission information of views into smart contracts. Shared data management should be conducted after the peer has been authorized. 

%Any modifications on the existing shared data should be updated to all sharing peers immediately. 

%We store the sharing peers' identity not pointers to the raw data into smart contracts in case of private data leakage. 

%Each sharing peer will receive the notification from smart contracts after updates on shared data are verified. Then each peer will request the newest data from updater and then use it to update his local full data. Smart contracts will refuse any further operations on shared data before all sharing peers have pulled the newest data to their local database.

In this paper, our contributions are as follows.
\begin{enumerate}

    \item We proposed an idea that a full medical record can be partitioned to multiple fine-grained data pieces shared with different peers and all data should reside on each user's local database.
    
    \item We solved the data synchronization between a full record and fine-grained data views after updates on a view by bidirectional transformation technology.
    
    \item We designed a decentralized medical data sharing architecture and applied blockchain to control permission for fine-grained data views.
        
   % \item We sketched procedures for data management (i.e., Create, Read, Update, Delete) operations on shared data.
    
%    \item We surveyed existing blockchain-based medical data sharing solutions and clarified their disadvantages, which are presented in Section \ref{related work};

\end{enumerate}

The remainder is organized like this. Section \ref{preli} gives some preliminaries about blockchain and bidirectional transformations. Section \ref{system} sketches our system design and provides an implementation architecture. Section  \ref{threats} discuss security threats of our system and potential countermeasures as our future work. Section \ref{related work} compares our work with existing ones to clarify our improvement over them. Section \ref{conclude} concludes and directs our future work.

\section{Preliminaries}
\label{preli}

    \subsection{Blockchain}
    Proposed with Bitcoin \cite{nakamoto2008bitcoin} in 2008, blockchain technology has been widely used in many fields. Blockchain provides a solution for data storage, data transfer and consensus protocol in a distributed and decentralized environment. Generally speaking, blockchain is a shared ledger and replicated by all nodes on a distributed network, which records the valid historical transactions in chronologically chained blocks. The nodes which generate new blocks by solving a computational puzzle (the proof-of-work problem) are called miners.
    
    Not only can support the platform of cryptocurrency, but blockchain can also be applied to other scenes. Ethereum \cite{wood2014ethereum} extends Bitcoin with a built-in Turing-Complete programming language so that one can use this scripts (i.e., Ethereum Virtual Machine (EVM) bytecodes) to write programs (i.e., smart contracts\footnote{Hyperledger and others still provide platforms to write smart contracts.}) on the blockchain. For example, we can write a smart contract in Solidity\footnote{https://solidity.readthedocs.io/en/v0.5.2/} and then compile it to EVM code. Besides the user accounts controlled by private keys like in Bitcoin, the accounts for smart contracts are allowed in Ethereum. Anyone can build decentralized applications which consist of a collection of smart contracts. Once a transaction involving smart contract creation gets confirmed, an address is generated for the contract and later anyone can send transactions to this address to execute the programs on it. A smart contract transaction is enforced when a miner includes it in a new produced block. Other nodes will validate it and re-run contracts if it is valid.
    
    \subsection{Bidirectional transformations}
    %Two directions are good, but hard to keep well behaveness. Once change one side, need to modify another side.
    % BXs: write one program to express two directions 
    % get-based may relate to multiple puts for one get
    % put-based is injective and better 
    Maintaining consistency between different data representations having overlapping contents is important \cite{abou2018introduction}. For example, in databases, a view table can be produced by querying a base source table; this view table can be modified, in which case we will want to ``restore consistency", i.e.,  we need to change the source such that the modified view coincides with the result of the query on the changed source --- this is the well-known view update problem \cite{bancilhon1981update}. To achieve this, one may consider providing two separate programs to represent the two directions to propagate updates from one side to the other. But it is hard to prove that the source and view can still be kept consistent after updates. Bidirectional transformations (BXs) were proposed  \cite{czarnecki2009bidirectional} to solve this.
    
    BX programs\footnote{The BXs we refer to in this paper are asymmetric lenses \cite{foster2007combinators}, one of the synchronization models studied by the BX community. } can be invoked in two ways as forward and backward transformations. A forward transformation (denoted as \emph{get}) extracts some information from the source to build an abstract view, and the backward transformation (denoted as \emph{put}\footnote{\emph{put} is not a simple inverse of \emph{get}. Instead, it accepts the view and the original source as input and produces an updated source as output.}) embeds information of the view back into the source and produces an updated source. This pair of transformations should satisfy the {\em round-tripping} laws (also referred to as {\em well-behaveness}) called \emph{PutGet} and \emph{GetPut}. 
    \begin{align}
    get (put(\textbf{source}, \textbf{view}))= \textbf{view} \tag {\emph{PutGet}} \\
    put (\textbf{source}, get (\textbf{source})) = \textbf{source} \tag {\emph{GetPut}} 
    \end{align}
    
    Intuitively, \emph{GetPut} states that no update should be performed on the source when there is no change on the view, while \emph{PutGet} hints that \emph{put} should take all updates on the view into account so that the view can be regenerated from the updated source by \emph{get}. 
    The most distinguished point of BX is that a view can contain only a part of a source. With respect to some consistency between a source and a view, BX programs can synchronize the source and view, and their well-behavedness guarantees that the source and view are kept consistent after updates on either side. There are some languages for constructing well-behaved BX programs, such as Boomerang\cite{bohannon2008boomerang}, BiGUL \cite{bigul} and HOBiT\cite{matsuda2018hobit}.
    
%    One BX program can represent the two transformations and we can achieve the transformation from one side to the other by invoking the  \emph{get} or \emph{put} direction of the program. 

    %which has been developed for putback-based bidirectional programming.

\section{System architecture}
\label{system}

In this section, we first describe a scenario of distributed medical data sharing among doctor, patient, and researcher in Section \ref{scenario} and propose a system architecture to model this scenario in Section \ref{architecture}. Then we present the solutions for synchronization between full medical record and views using BXs and access control on views applying blockchain in Section \ref{system design}. Our system can not only allow updates on shared data but other operations such as creating data, which are described in Section \ref{dataManage} and explained by a concrete case in Section \ref{updateCase}. 

%     Part A describes the data\footnote{In our prototype, medical data are entered directly by users. Later we may consider using the data from wearable devices.} distribution between sharing peers. Part B explains our system architecture.

%To simplify our expression next, we adopt nodes and users to denote devices connecting to blockchain network and stakeholders (doctors, patients, etc.) in medical scenarios respectively. Users sharing data are called sharing peers.
     
\subsection{Scenario description}
\label{scenario}

\begin{figure}[htbp]
    \centerline{\includegraphics[width=250pt]{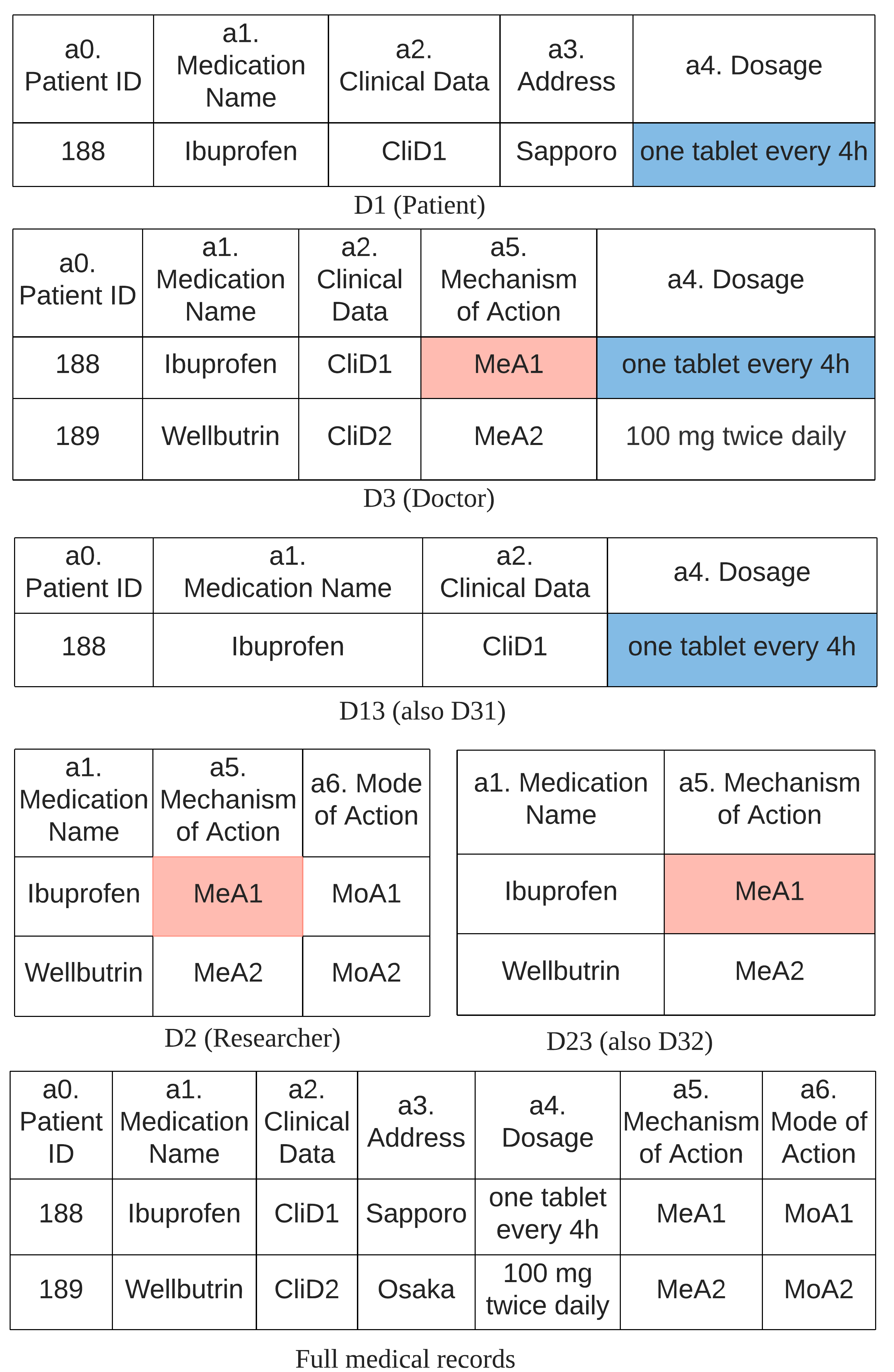}}
    \caption{Data distribution}
    \label{dataRepresentation}
\end{figure}

%A user can share different views with different peers based on the base (source) table, as shown in Fig. \ref{workflow}.  
Table "Full medical records" in Figure \ref{dataRepresentation} shows patients' full medical records which includes seven attributes: a0. patient ID, a1. medication name, a2. clinical Data, a3. address, a4. dosage, a5. mechanism of action, a6. mode of action. Suppose there are three users: Patient, Researcher, and Doctor. Each user only stores some attributes of full records on their local databases. For instance, Patient only keeps the attributes from a0 to a4 on Table D1 and Researcher retains attribute a1, a5 and a6 on Table D2. Doctor manages attributes a0-a2, a4, and a5 on Table D3 and generates two view tables D31 and D32 by querying special attributes from D3. D31 and D32 are used to be shared with Patient and Researcher respectively. Note that D13 and D31 are identical tables but D13 is stored in Patient's database and D31 resides in Doctor's database. Similarly, attributes a1 and a5 are shared between Researcher and Doctor, which are shown in the view Table D23 on Researcher's database and Table D32 on Doctor's storage.  We note that the formats and contents of shared data are predefined by sharing peers. 

 %D32 and D23 should contain the same contents, which means either one is updated and the other one need be modified to become identical with it again. Similarly, Patient and Doctor share the same contents that stored in D13 and D31 separately in their sides. 
 
\subsection{System architecture}
\label{architecture}

Our system architecture is shown in Fig. \ref{systemArchitecture}, which contains following components:

\begin{figure}[htbp]
    \centerline{\includegraphics[width=250pt,height=200pt]{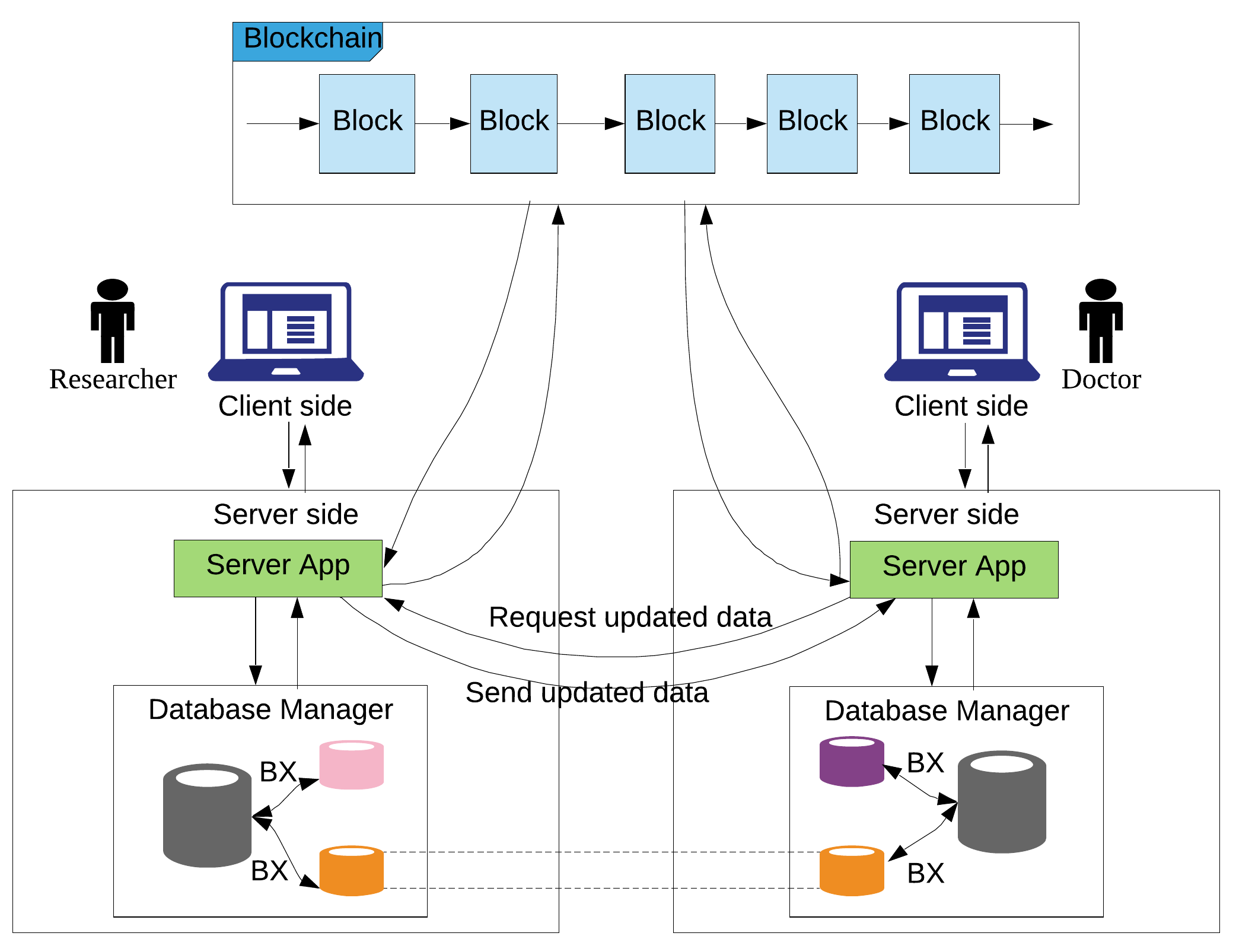}}
    \caption{System Architecture}
    \label{systemArchitecture}
\end{figure}

\begin{itemize}
    \item \emph{Client side} controls the interaction between users and other components.
    
    \item \emph{Server App} acts as a mediator to interact with other components. 
    
    \item \emph{Blockchain} keeps the manage permission of shared data on smart contracts and notifies sharing peers the change on them.
    
    %     Also, front-end user interfaces communicate with blockchain network via nodes. The smart contract on blockchain contains the metadata of shared medical data and maintain a management log that store historical modifications for metadata.
    \item \emph{Database manager} disposes of the synchronization between shared data and local data according to consistency logic relations. These synchronizations are implemented by executing BX programs.
    
    \item \emph{Database}: each user has a full database and many data pieces shared with other users. The latter (seen as a view) can always be reproduced from the former (seen as a source).
    
\end{itemize}

Now we discuss details about this architecture.

% \item nodes can split total data into multiple pieces which can be shared with different nodes, which make sharing exists among only a group of nodes. So that it can avoid the interference or attack from other nodes. 

%    \item Meanwhile, data provider can choose what they want to share with others without exposing sensitive or private information.

%    \item Permission to updates to medical data are stored in smart contracts so that blockchain can prevent operations from malicious nodes. 

Firstly, medical data always stay in each peer's local database and data transfer only exists between sharing peers, which avoid data being leaked to the third party so as to keep shared data security. The data can not only be provided by doctors. Instead, each node can be a shared data provider. As referred in \cite{chung2018using},  many clinics encourage patients to collect data by themselves that are supposed to be gathered by doctors and expect to increase clinic efficiency and promote patient awareness. 

Moreover, blockchain's consensus protocol scheme keeps the shared data between sharing peers the same after updates since each peer will receive the notification from contracts and request new shared data from other sharing peers. Additionally, any modification on shared data can be recorded on the blockchain. Blockchain properties such as immutability, auditablility, and transparency enable nodes to check and review update history on shared data. Still, simultaneously updates to the same shared data by multiple peers are forbidden. Smart contracts dispose of the updates according to received requests in chronological order. If a transaction for updates on shared data has been included in a block, then other requests on this shared data will not be accepted, i.e., one block can contain one transaction at most on some shared data at one time. This can promise that only when all sharing peers have had the newest shared data can they execute further operations.

%Lastly, this system architecture can also be applied to other data sharing scenarios.

%    \item Any update on shared data can be reflected in local total database by using \emph{put}. Consistency between shared data and local total data are firmly promised by BX. 

\subsection{System design}
\label{system design}
%\subsection{Data distribution}

\subsubsection{Synchronization between source and views}
\label{synchronization}

% If we build a big contract which is a list of list and each small list is a contract for a shared data

%\footnote{In our tentative prototype, medical data are entered directly by users. Later we may consider using the data from wearable devices.} 

%which means that D13 may have different representations with D32.

%For each table, there are some attributes such as medication name and address on D1.  
As we said before, BX programs are used to synchronize the full data and shared data. Shared data can be seen as views which can be produced from full records named as sources. For example in Fig. \ref{dataRepresentation}, Table D13 is shared by Patient and Doctor and can be produced from D1 by \emph{$BX_{13}$-get} (i.e., applying the get direction of the BX program between D1 and D13). If D13 is modified, then D1 need to be updated from original D1 and D13 by using  \emph{$BX_{13}$-put}(i.e., invoking put the direction of the BX program between D1 and D13) to ensure that the modified D13 can be regenerated from the updated D1.  

In our design, shared data between any two peers are not exposed to the third party, which can keep data privacy between them to some degree. For example, any operations on D23 or D32 can only be known by Researcher and Doctor and Patient has no information about this. However, If D31 and D32 have some overlapping data, after D32 is modified, D31 might need to be regenerated by using \emph{$BX_{31}$-get} on modified D3 which can be produced by applying \emph{$BX_{32}$-put}.

%\subsection{Case analysis}

%
%Dependency 
%Incremental update
%Dependency graph
%
%Just send them data
%Put online; give them link
%Give data to each person
%
%Advantages of having data Locally not on server: 
%fast/easy to access
%Full control own data; change format
%exposed to others (since public available; sensitive data)
%scalability (multi-access server; slow)

\subsubsection{Access control on views}
\label{permission}

\begin{figure}[htbp]
    \centerline{\includegraphics[width=250pt,height=150pt]{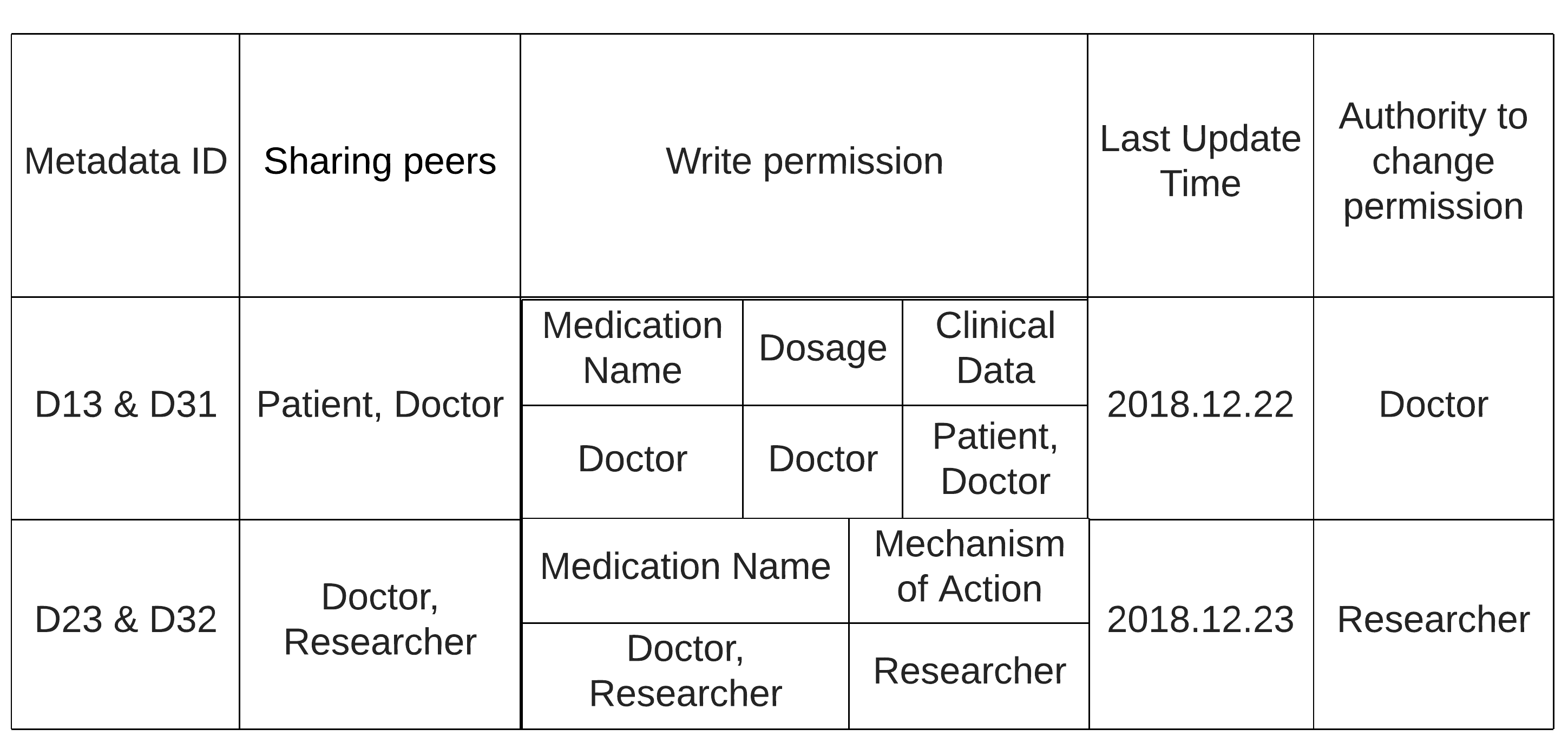}}
    \caption{Metadata collection in smart contract}
    \label{metadata}
\end{figure}

Figure \ref{metadata} presents a metadata collection table which dictates the update permission on each attribute of the shared data. These kinds of tables reside in smart contracts on the blockchain. Each metadata entry corresponds to a shared table. For example, the entry for D13 or D31 declares that it is shared by Patient and Doctor and Doctor can update all attributes value but Patient can only change the clinical data. The ``Latest Update Time" shows when the metadata was modified most recently. The value on the attribute ``Authority to Change Permission"  indicates the user who can modify other users' permission. For instance,  because D13 and D31 are initially produced by Doctor and shared with Patient and then Doctor can change Patient's access permission. Doctor can change the permission for updating ``Dosage" from  ``Doctor" to ``Doctor, Patient" so that Patient can also update the  ``Dosage" later.

If users want to share data, they need to form an agreement on the structure of the shared table and register the corresponding metadata on smart contracts. Suppose Doctor initiates the data sharing with Patient. According to their agreement, he will deploy a smart contract on blockchain which stipulates the metadata about the shared data, such as sharing peers (i.e., Patient and Doctor) and so on. 

%the ability to modify the update permission are encoded in the field named as ``authority to change permission". 

%Since smart contract cannot be altered after it is deployed on blockchain. 
%
%There are two ways to change the permission for updates to shared data:
%\begin{enumerate}
%    \item deploy a new contract and notify all nodes on blockchain that this new one should be used to control permission;
%    \item update the state of variables in contract.
%\end{enumerate}
%We choose the latter way. In Fig. \ref{dataRepresentation}, 

\subsection{Data Management}
\label{dataManage}

\begin{figure}[htbp]
    \centerline{\includegraphics[width=250pt,height=170pt]{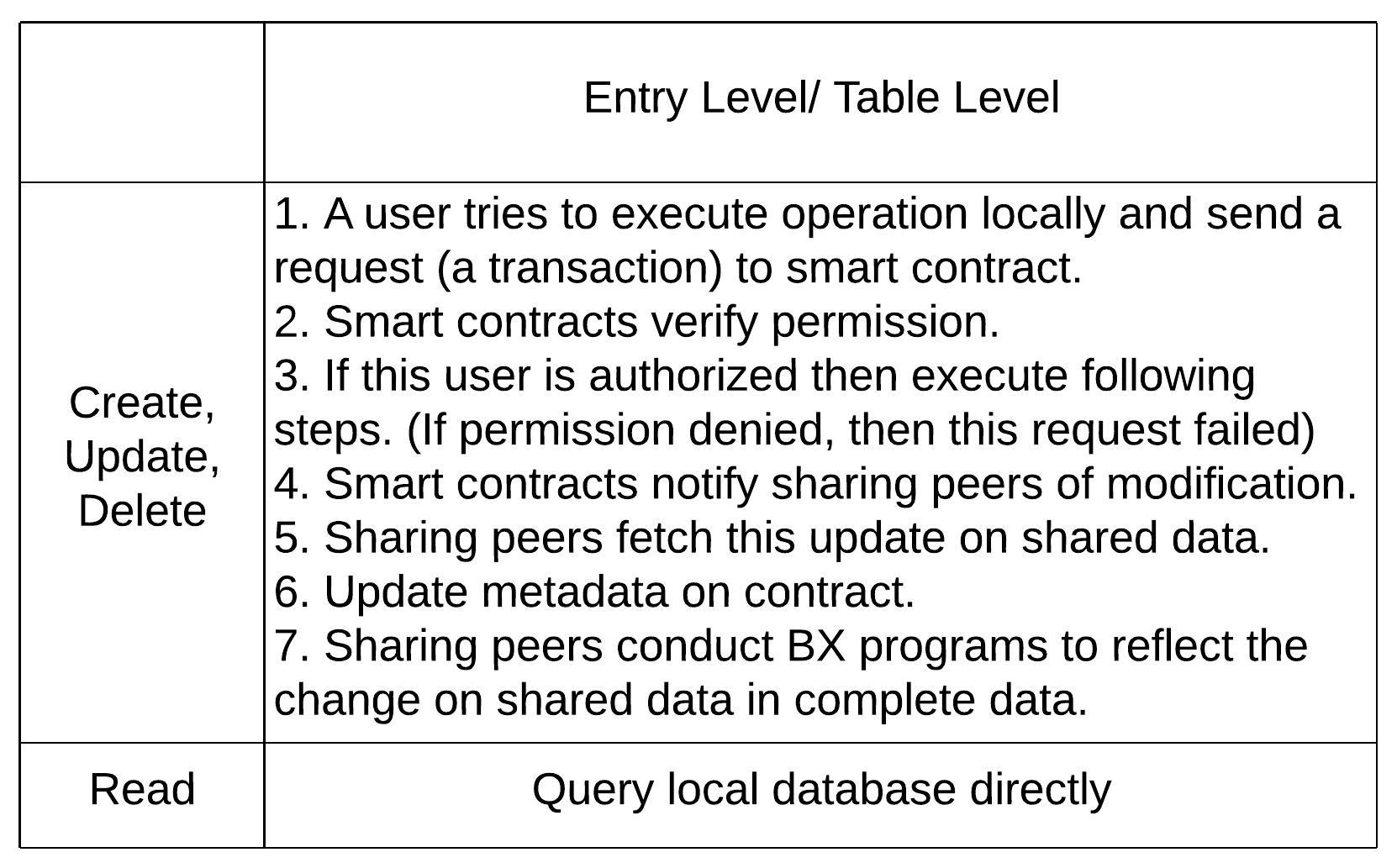}}
    \caption{CRUD operations on shared data}
    \label{data manage}
\end{figure}

In Fig. \ref{data manage}, we briefly sketch the procedures for CRUD  (i.e., Create, Read, Update, Delete) operations on shared data considering entry level or table level. For Read operation, since shared data stay in users' local databases, they can just execute a query to get the shared data. We describe the workflow for updating an item on Section \ref{updateCase}.

\subsection{Case analysis for updating shared data}
\label{updateCase}

\begin{figure}[htbp]
    \centerline{\includegraphics[width=270pt,height=220pt]{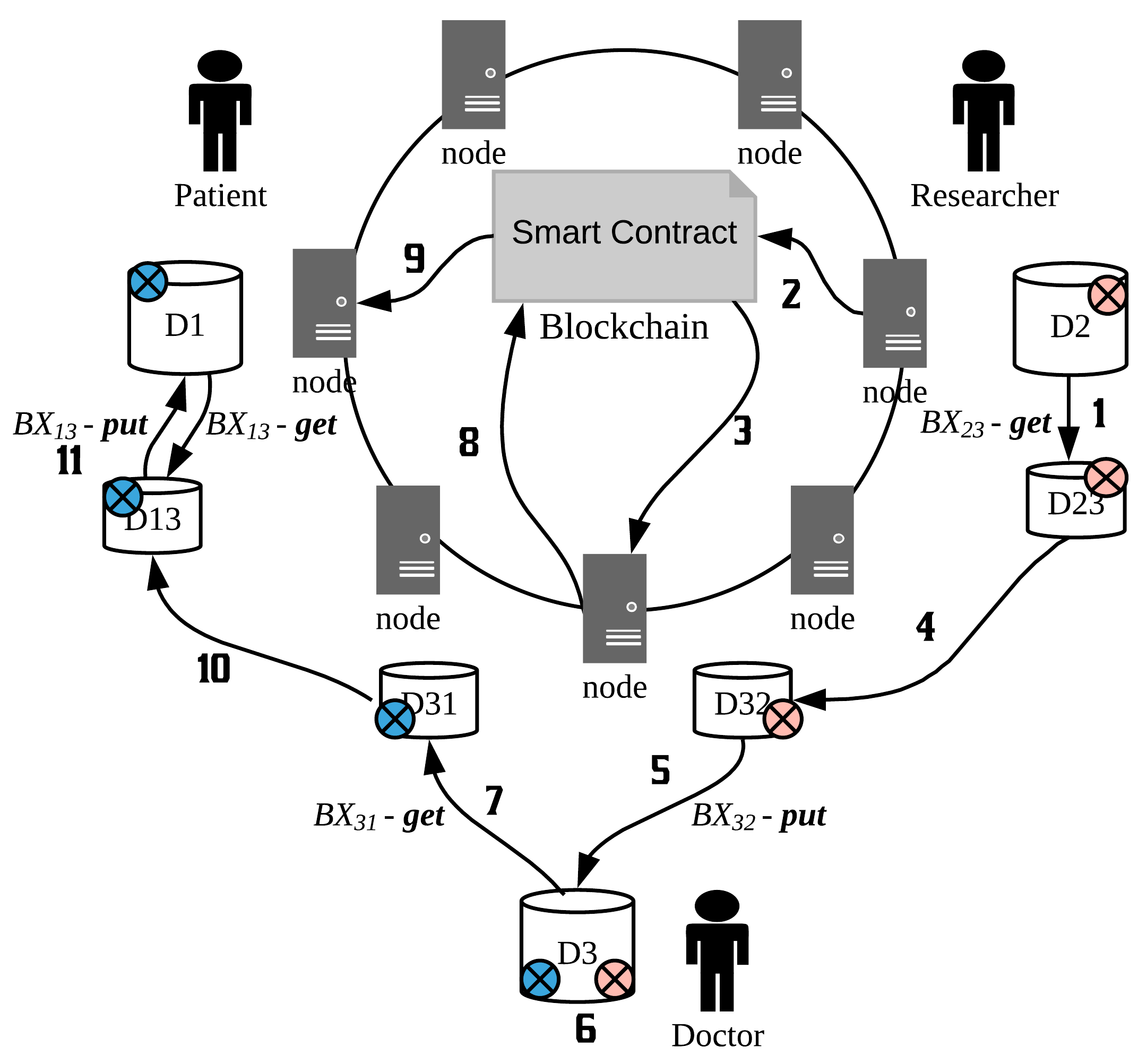}}
    \caption{A workflow for updating data fields of shared data}
    \label{workflow}
\end{figure}

Figure \ref{workflow} depicts a scenario where the researcher initiates the update the shared data. We use notations in Fig. \ref{dataRepresentation}. The numbers indicate the corresponding operations sequence. The red and blue circles indicate the updated places. Steps 1 - 5 and Steps 7 - 11 perform procedures for an operation on Fig. \ref{data manage}. Notably, Step 6 checks whether D32 and D31 have some dependencies since they might have some overlapping data. Our work leaves the initialization of shared data to future work and we only consider management on existing shared data. 

Let us try to describe this workflow using the data in Fig.~ \ref{dataRepresentation}. After updating the ``MeA1" on D2, the Researcher wants to propagate the update to the shared data D23 so he uses the \emph{$BX_{23}$-get} to regenerate D23 (Step 1). Then he will call a smart contract via a trusted node connected to blockchain by sending the request for updates to the D23 (Step 2). Note that the smart contract in Fig. \ref{metadata}  records all permission info about that D23. 
The smart contract will be executed until all nodes form a consensus on this update request, which means Researcher is permitted to update D23. Each node will conduct the smart contract locally. The entry related D23 \& D32 of the contract is modified and the Doctor will receive notification that D32 needs to be modified (Step 3). Then he will ask for data from Researcher by sending a data request message and use the latest shared data to refresh D32 (Step 4). After that, Doctor will use \emph{$BX_{32}$-put} to reflect the change on D32 in D3, i.e., update the ``MeA1" to a new name (Step 5). Since Doctor shared some data (D31) with Patient, he needs to check whether D31 needs to be reproduced (Step 6). (If there is no need to reproduce, Step 6 - 11 will not be performed.) For example, Doctor may want to modify the ``Dosage" on D31. He will use \emph{$BX_{31}$-get} to regenerate D31 (Step 7) and request smart contract for permission to update D13 (Step 8). Once allowed, Patient will receive a notification about the change on ``Dosage" (D31) (Step 9) and ask Doctor to send the updated D31. After Patient gets the modified D31 (Step 10), he will use this to update D1 via \emph{$BX_{13}$-put} (Step 11).

\section{Threats and countermeasures}
\label{threats}
 In this section, we identify some threats to our system and propose relating countermeasures.
\subsubsection{Throughput}
We might employ smart contracts to control access to shared data. Usually, the block creation time is approximately 12 seconds on Ethereum. We argue that this time interval is acceptable since nodes may choose to collect a lot of updates and then send requests to contracts. It is not so urgent for a patient or doctor to get the immediate updated shared data.

\subsubsection{Correctness of smart contracts}
Smart contracts might be inconsistent with specifications. We may apply some theorem prover such as Coq\cite{huet2004coq} to verify the correctness of smart contracts to prevent these attacks.

\subsubsection{Public blockchain}
Once deployed to the public Ethereum blockchain, transactions related to our systems might not be chosen into a block by miners. So a private blockchain might be a better choice for our system.

\subsubsection{Incentive}
Like in \cite{dagher2018ancile}, we do not include any incentive for mining beyond the use of our system. We presume that all nodes on the blockchain already have incentives to keep medical data from being illegal access or updates.

\section{Related work}
\label{related work}
In this section, we review existing blockchain-based research on the medical data sharing field and list the advantages of our system compared with them.

%Zyskind et al. suggested using blockchain for access control in \cite{zyskind2015decentralizing} where encrypted data reside on the third party storage. But data might be exposed by this ``trusted" the third party so that data privacy is violated.

The idea of introducing Blockchain technology to healthcare was presented firstly in \cite{yue2016healthcare} where they use blockchain for data storage to guarantee that medical data cannot be modified by anyone. Also, they designed a Healthcare Data Gateway (HDG) to control access to the shared data. However, the medical data size can become huge so that the data become burdens for blockchain nodes' storage since each node has the same copy of blockchain. Usually, the size of metadata is smaller than data. (It also depends on the structure of metadata and data.) We store metadata on smart contracts so as to reduce the storage pressure for each blockchain node.
%Similarly, Patientory \cite{mcfarlane2017patientory} proposed that the medical data is stored directly on the HIPPA-compliant blockchain database.  

MedRec \cite{azaria2016medrec} chose to store raw medical data on providers' database and patients can download the data from it after authorized by smart contracts on the blockchain. They aimed to enable patients to engage in their healthcare. Whereas in our system, all parties, such as doctors, patients, and researchers can benefit from sharing data with others. MedRec recognized that not all provider data such as physician intellectual property can be exposed to patients \cite{us2017individuals, grossman2011clinical} so that they do not claim to manage contents automatically from physician's output. Instead, we allow each node to share a piece of medical data not total but still keep consistency between them after the updates to the shared ones. Additionally, any modifications on data shared by two nodes will not be disclosed to the third party which keeps the consistency only exists in sharing peers. Moreover, since all shared data with others can be a part of each nodes' local total databases, we can decide whether one shared data have some influence on the other shared pieces and then propagate this change to the third party.

Dubovitskaya et al. gave an architecture for managing and sharing medical data for cancer patient care \cite{dubovitskaya2017secure}. They stored encrypted categorized shared data on the cloud and related metadata in blockchain and implemented the prototype on Hyperledger\cite{hyperledger2017hyperledger}. The access control policy is defined in the chaincode Logic by patients. Whereas we think that each data provider, not just patients can use smart contracts to encode the control policy when they deploy them to blockchain. % Ethereum. 

Notably, these three solutions and others \cite{liu2018bpds,xia2017bbds,amofa2018blockchain,dagher2018ancile,fan2018medblock} mostly targeted the access problem on shared data but did not pay much attention to updates on them. Additionally, they presumed that different parties can share the same data. Unlike them, we aim to solve the update issues on the shared data and allow one party to split full data (i.e., source) into multiple pieces (i.e, views) which are shared with different parties but still keep consistency between source and views.

%Distributed Authorized Medical Data Management
\section{Conclusion}
\label{conclude}

Medical data sharing is necessary and important, which allows stakeholders on medical scenarios to contribute their knowledge to better the medical treatment. Users may have different interests in the same full medical record. Some peers might update some values of fields in the existing data. These updates need to be propagated to sharing peers. Our architecture divides a record into pieces that shared with different users separately, which can protect data privacy by limiting essential data between two peers and reduce the unrelated data interference. Any updates on data pieces can be synchronized to full records by bidirectional transformations. Moreover, based on smart contracts on the blockchain, we can promise that only authorized users can update the existing shared data and only when all peers have updated to the latest data contents can they continue the operations on shared data.   

We are still developing the prototype to implement our idea.
In the future, we will use real patient data to do experiments but use some de-identification technology to protect patient data from being exposed. 

%\section*{Acknowledgment}
%The authors would like to thank all PRL lab members for their helpful advice in the paper draft.  Thanks Hsiang-Shang Ko for revising the BX parts and providing valuable suggestions to our work. Thanks Van-Dang Tran's help on building the initial data distribution idea sketch. We would like to thank anonymous reviewers for reviewing this paper. This work is partially supported by the Japan Society for the Promotion of Science (JSPS) Grant-in-Aid for Scientific Research (S) No. 17H06099 and Scientific Research (A) No. 18H04093.

\bibliographystyle{IEEEtran}
\bibliography{ref}

\end{document}